\documentstyle[preprint,aps]{revtex}
\tighten
\draft
\widetext
\preprint{YITP-00-81}
\bigskip
\bigskip
\begin{document}
\title{Time-like Extra Dimensions without Tachyons or Ghosts}
\medskip
\author{Alberto Iglesias\footnote{E-mail:
iglesias@insti.physics.sunysb.edu} and Zurab Kakushadze\footnote{E-mail: 
zurab@insti.physics.sunysb.edu}}
\bigskip
\address{C.N. Yang Institute for Theoretical Physics\\ 
State University of New York, Stony Brook, NY 11794}

\date{December 4, 2000}
\bigskip
\medskip
\maketitle

\begin{abstract} 
{}Some time ago Dvali, Gabadadze and Senjanovi{\'c} \cite{DGS} 
discussed brane world scenarios with time-like extra dimensions.
In this paper 
we construct a solitonic 3-brane solution in the 5-dimensional 
Einstein-Hilbert-Gauss-Bonnet theory with the space-time signature
$(-,+,+,+,-)$. The direction transverse to the brane is the second time-like
direction. The solitonic brane
is $\delta$-function like, and has the property that gravity is completely
localized on the brane. That is, there are no propagating degrees of freedom
in the bulk, while on the brane we have purely 4-dimensional Einstein gravity.
In particular, there are no propagating tachyonic or negative norm states
even though the extra dimension is time-like.  
\end{abstract}

\section{Introduction}

{}In the Brane World scenario the Standard Model gauge and matter fields
are assumed to be localized on  
branes (or an intersection thereof), while gravity lives in a larger
dimensional bulk of space-time 
\cite{early,BK,polchi,witt,lyk,shif,TeV,dienes,3gen,anto,ST,BW,Gog,RS,DGP,DG,zura}. Usually it is assumed that the extra dimensions transverse to the branes
are space-like. This is because otherwise we generically expect difficulties
with propagating tachyonic and/or negative norm states whose appearance
is due to the presence of more
than one time-like directions. Various issues arising in brane world scenarios
with time-like extra dimensions were discussed in \cite{DGS} 
(for subsequent developments, see \cite{timelike}).  

{}In this paper we would like to ask whether
the aforementioned difficulties can be 
avoided in brane world scenarios with time-like extra dimensions.
As we will argue in the following,
the answer to this question appears to be positive. Thus, recently in 
\cite{alberto} we constructed a
(flat) {\em solitonic} codimension-one brane world solution (with
a space-like extra dimension), where gravity is 
completely localized on the brane. That is, the graviton
propagator in the bulk {\em vanishes}, while it is non-trivial on the
brane. In this paper we point out that such a solution also exists in the case
where the extra dimension is {\em time-like}. In this solution we also have
completely localized gravity, and we have no propagating
tachyonic or negative norm states as there are no propagating degrees of 
freedom in the bulk. Moreover, just 
as in the solution of \cite{alberto}, albeit the
classical background is 5-dimensional, the quantum theory (at least 
perturbatively) is actually 4-dimensional. In particular, there are no loop
corrections in the bulk.

{}The setup within which we construct this solitonic brane world solution
is the 5-dimensional Einstein-Hilbert theory with a (positive) cosmological
term augmented with a Gauss-Bonnet coupling (the signature of the 5-dimensional
space is $(-,+,+,+,-)$). The solitonic brane world
solution arises in this theory for a special value of the Gauss-Bonnet
coupling. The fact that there are no propagating degrees of freedom in the
bulk is then due to a perfect cancellation between the 
corresponding contributions coming from the Einstein-Hilbert and
Gauss-Bonnet terms.
Since the bulk theory does not receive loop corrections, the
classical choice of parameters such as the special value of the
Gauss-Bonnet coupling (or the Gauss-Bonnet combination itself) does {\em not}
require order-by-order fine-tuning. Also, we can embed
this solution in the (minimally) supersymmetric setup, 
where we still have no propagating degrees of
freedom in the bulk, while on the brane we have completely localized 
{\em supergravity}. Here the solitonic brane is a BPS solution (with
vanishing brane cosmological constant), which preserves 1/2 of the original
supersymmetries.

\section{The Setup}

{}In this section we discuss the setup within which we will discuss the
aforementioned solitonic brane world solution. The action for this model
is given by (for calculational convenience we will keep the number of 
space-time dimensions $D$ unspecified):
\begin{eqnarray}
 S= &&M_P^{D-2}
 \int d^D x \sqrt{-G} \Big(R-\Lambda +\nonumber\\
 \label{actionGB}
 &&\lambda\left[R^2-4R_{MN}^2+R_{MNST}^2\right] \Big)~,
\end{eqnarray}
where $M_P$ is the $D$-dimensional (reduced) Planck scale, and 
the Gauss-Bonnet coupling $\lambda$ has dimension $({\rm length})^2$.
Finally, the bulk vacuum energy density $\Lambda$ is a constant.
The $D$-dimensional space-time has signature $(-,+,\dots,+,-)$.

{}In the following we will be interested in solutions 
to the equations of motion following from the action (\ref{actionGB}) 
with the warped \cite{Visser} metric of the form 
\begin{equation}\label{warped}
 ds_D^2=\exp(2A)\eta_{MN} dx^M dx^N~,
\end{equation}
where $\eta_{MN}\equiv{\rm diag}(-1,+1,\dots,+1,-1)$,
and the warp factor $A$, which is a function of $z\equiv x^D$, 
is independent of the other $(D-1)$ coordinates $x^\mu$.
With this Ansatz, we have the following
equations of motion for $A$ (prime denotes derivative w.r.t. $z$):
\begin{eqnarray}
 \label{A'GB}
 &&(D-1)(D-2)(A^\prime)^2\left[1+\kappa\right]- 2\Lambda \exp(2A)=0~,\\
 \label{A''GB}
 &&(D-2)\left[A^{\prime\prime}-(A^\prime)^2\right]
 \kappa=0~,
\end{eqnarray}
where
\begin{equation}\label{factor1}
 \kappa\equiv 1+2(D-3)(D-4)\lambda(A^\prime)^2\exp(-2A)~.
\end{equation}
For generic values of $\Lambda$ and $\lambda$ such that the above system of 
equations has a solution, the corresponding background is expected to suffer
from propagating negative norm states as we have two time-like 
directions.

{}There, however, also exists a solution which is free of propagating negative
norm states. Thus, consider the case where
\begin{equation}\label{fine}
 \Lambda=- {{(D-1)(D-2)}\over{(D-3)(D-4)}}{1\over 4\lambda}
\end{equation}
with $\lambda<0$ and $\Lambda>0$. Then we have the following solution 
(we have chosen the integration constant such that $A(0)=0$):
\begin{equation}\label{solutionGB}
 A(z)=-\ln\left[{|z|\over\Delta}+1\right]~,
\end{equation}
where $\Delta$ is given by
\begin{equation}\label{Delta}
 \Delta^2= -2(D-3)(D-4)\lambda~.
\end{equation}
Note that $\Delta$ can be positive or negative. In the former case the
volume of the $z$ direction is finite: $v=2\Delta/(D-1)$. 
On the other hand, in the latter case it is infinite.
As we will see in the following, the negative $\Delta$ case corresponds to
a non-unitary theory. 

{}Note that $A^\prime$ is discontinuous at $z=0$, and $A^{\prime\prime}$
has a $\delta$-function-like behavior at $z=0$. Note, however, that
(\ref{A''GB}) is still satisfied as in this solution
\begin{equation}\label{factor}
 \kappa=0~.
\end{equation}
Thus, this solution
describes a codimension one {\em soliton} 
with a time-like transverse dimension. 
The tension of this soliton,
which is given by
\begin{equation}
 f=-{4(D-2)\over \Delta}M_P^{D-2}~,
\end{equation}
is negative for $\Delta>0$, and it is positive for $\Delta<0$. The
aforementioned non-unitarity in the latter case is, in fact, attributed to
the positivity of the brane tension (but has nothing to do with the fact that
we have two time-like dimensions). Note that this is opposite to the case with
a space-like extra dimension \cite{alberto}.
Here and in the following we refer to the
$z=0$ hypersurface, call it $\Sigma$, as the brane. Note that the background
metric on the brane is the flat $(D-1)$-dimensional Minkowski metric
$\eta_{\mu\nu}={\rm diag}(-1,+1,\dots,+1)$.

\section{Gravity in the Solitonic Brane World}

{}In this section we would like to study gravity in the solitonic brane
world solution discussed in the previous section along the lines of 
\cite{alberto}. Let us study small fluctuations around the solution:
\begin{equation}\label{fluctu}
 G_{MN}=\exp(2A)\left[\eta_{MN}+{\widetilde h}_{MN}\right]~,
\end{equation}
where for convenience reasons we have chosen to work with 
${\widetilde h}_{MN}$
instead of metric fluctuations $h_{MN}=\exp(2A){\widetilde h}_{MN}$. 

{}Let us assume that we have matter localized on the brane, and 
let the corresponding conserved energy-momentum tensor be $T_{\mu\nu}$:
\begin{equation}\label{conserved}
 \partial^\mu T_{\mu\nu}=0~.
\end{equation}
The graviton field ${\widetilde h}_{\mu\nu}$ couples to $T_{\mu\nu}$ via
the following term in the action (note that ${\widetilde h}_{\mu\nu}=
h_{\mu\nu}$ at $z=0$ as we have set $A(0)=0$):
\begin{equation}\label{int}
 S_{\rm {\small int}}={1\over 2} \int_\Sigma d^{D-1} x ~T_{\mu\nu}
 {\widetilde h}^{\mu\nu}~.
\end{equation} 
In the following we will use the following notations for the component
fields:
\begin{equation}
 H_{\mu\nu}\equiv{\widetilde h}_{\mu\nu}~,~~~A_\mu\equiv
 {\widetilde h}_{\mu D}~,~~~\rho\equiv{\widetilde h}_{DD}~.
\end{equation}
 The linearized equations of motion for the component fields $H_{\mu\nu}$,
$A_\mu$ and $\rho$ read:
\begin{eqnarray}\label{EOM1oGB}
 &&\kappa\Big(\Omega_{\mu\nu}-\Sigma_{\mu\nu}^{\prime}-
 (D-2) A^\prime \Sigma_{\mu\nu}-\partial_\mu\partial_\nu\rho+\eta_{\mu\nu}
 \partial_\sigma\partial^\sigma 
 \rho+\nonumber\\
 &&\eta_{\mu\nu}\left[(D-2)A^\prime\rho^\prime
 +(D-1)(D-2)(A^\prime)^2\rho\right]\Big)+\nonumber\\
 &&4(D-4)\lambda\left[A^{\prime\prime}-(A^\prime)^2\right]e^{-2A}
 \Big(\Omega_{\mu\nu}-
 (D-3)A^\prime\Sigma_{\mu\nu}\Big)+\nonumber\\
 && 2(D-2)
 \left[2\kappa-1\right]\left[A^{\prime\prime}-(A^\prime)^2\right]
 \eta_{\mu\nu}\rho=\nonumber\\
 &&-M_P^{2-D} T_{\mu\nu} \delta(z)~,\\ 
 \label{EOM2oGB} 
 &&\kappa\Big(Q_\nu^\prime
 -\partial^\mu F_{\mu\nu}-(D-2)A^\prime \partial_\nu\rho\Big)=0~,\\
 \label{EOM3oGB}
 &&\kappa
 \Big(\partial^\nu Q_\nu +
 A^\prime \Sigma -
 (D-1)(D-2)(A^\prime)^2\rho\Big)=0~,
\end{eqnarray}
where $F_{\mu\nu}\equiv\partial_\mu A_\nu - \partial_\nu A_\mu$ 
is the $U(1)$ field strength for the graviphoton, 
\begin{eqnarray}
 \Omega_{\mu\nu}\equiv&&\partial_\sigma\partial^\sigma 
 H_{\mu\nu} +\partial_\mu\partial_\nu
 H-\partial_\mu \partial^\sigma H_{\sigma\nu}-
 \partial_\nu \partial^\sigma H_{\sigma\mu}-\nonumber\\
 &&\eta_{\mu\nu}
 \left[\partial_\sigma\partial^\sigma H-\partial^\sigma\partial^\rho
 H_{\sigma\rho}\right]~,\\
 \Sigma_{\mu\nu}\equiv &&H_{\mu\nu}^\prime-\partial_\mu A_\nu-
 \partial_\nu A_\mu -\eta_{\mu\nu}\left[H^\prime-2\partial^\sigma A_\sigma
 \right]~,\\
 Q_\nu\equiv &&\partial^\mu H_{\mu\nu}-\partial_\nu H~, 
\end{eqnarray}
while $H\equiv {H_\mu}^\mu$, and $\Sigma \equiv {\Sigma_\mu}^\mu$.

{}The above equations of motion are invariant under certain gauge
transformations corresponding to unbroken diffeomorphisms.  
In terms of the component fields $H_{\mu\nu}$, $A_\mu$ and $\rho$, the
full $D$-dimensional diffeomorphisms read:
\begin{eqnarray}\label{diff1}
 &&\delta H_{\mu\nu}=\partial_\mu{\widetilde\xi}_\nu+\partial_\nu{\widetilde
 \xi}_\mu-2\eta_{\mu\nu}A^\prime\omega~,\\
 \label{diff2}
 &&\delta A_\mu={\widetilde\xi_\mu}^\prime +\partial_\mu \omega~,\\
 \label{diff3}
 &&\delta\rho=2\omega^\prime+2A^\prime\omega~,
\end{eqnarray}
where $\omega\equiv {\widetilde \xi}_D$.
It is not difficult to check that
the equations of motion (\ref{EOM1oGB}), (\ref{EOM2oGB}) and (\ref{EOM3oGB})
are invariant under these full $D$-dimensional diffeomorphisms. That is, there
are no restrictions on $\omega$ or ${\widetilde \xi}_\mu$ or derivatives 
thereof including at $z=0$. In particular, this is the case for the solitonic
brane world solution despite its $\delta$-function-like structure. The reason 
for this is that this solution being a soliton 
does not break the full $D$-dimensional 
diffeomorphisms explicitly but only {\em spontaneously}.

{}Since we have the full $D$-dimensional diffeomorphisms, we can always gauge
$A_\mu$ and $\rho$ away. In fact, in the following we will see that for the
solitonic brane world background this can indeed be done without introducing 
any inconsistencies. However, before we adapt this gauge fixing, 
we would like to make the following 
important observation. Note that for the solitonic brane world solution
(\ref{solutionGB}) with $\Delta$ given by (\ref{Delta}) we have (\ref{factor}).
On the other hand, this vanishing factor $\kappa$
is precisely the one that multiplies
the terms in (\ref{EOM1oGB}), (\ref{EOM2oGB}) and (\ref{EOM3oGB}) 
corresponding to the propagation of the fields $H_{\mu\nu}$, $A_\mu$ and
$\rho$ in the bulk. That is, in the solitonic brane world solution these fields
do {\em not} propagate in the time-like 
$z$ direction at all. This is due to a cancellation
between contributions of the Einstein-Hilbert and Gauss-Bonnet terms into
the bulk propagator in this background.
On the other hand, (some of)
these fields do propagate on the brane. Indeed, in the above background we
have 
\begin{equation}
 A^{\prime\prime}-(A^\prime)^2=-{2\over \Delta} \delta(z)~.
\end{equation}
Then (\ref{EOM1oGB}) gives the following equation of motion (note that
(\ref{EOM2oGB}) and (\ref{EOM3oGB}) are trivially satisfied in this 
background):
\begin{eqnarray}\label{EOM1oGBX}
 &&\Big(\Omega_{\mu\nu}-(D-3)A^\prime \Sigma_{\mu\nu}-
 {(D-2)(D-3)\over \Delta^2}\eta_{\mu\nu}\rho\Big)\delta(z)=\nonumber\\
 &&-{\widehat M}_P^{3-D} 
 T_{\mu\nu} \delta(z)~,
\end{eqnarray}
where
\begin{equation}
 {\widehat M}_P^{D-3}\equiv {4\Delta\over{D-3}}M_P^{D-2}~,
\end{equation}
and in the following we will identify ${\widehat M}_P$ with the 
$(D-1)$-dimensional Planck scale.

{}Thus, as we see, in the negative tension solution there are no propagating
tachyonic or
negative norm states in the bulk or on the brane, and the theory is unitary.
Moreover, since we have no propagating degrees of freedom in the bulk, we
have no loop corrections in the bulk either. This implies that, albeit the 
classical background is $D$-dimensional, the quantum theory (at least 
perturbatively) is actually $(D-1)$-dimensional. In particular, the condition
(\ref{fine}) is stable against loop corrections. 

\subsection{Completely Localized Gravity}

{}Next, we would like to see what is the solution to the equation of
motion (\ref{EOM1oGBX}). First, note that, as we have already mentioned,
we can always gauge $A_\mu$ and $\rho$ away. That is, these fields are {\em
not} propagating degrees of freedom. Note that after this gauge fixing the
residual gauge symmetry is given by the $(D-1)$-dimensional diffeomorphisms
for which $\omega\equiv 0$, and ${\widetilde \xi}_\mu$ are independent of $z$.
Second, note that the second term in parentheses on the l.h.s. of
(\ref{EOM1oGBX}) contains 
$A^\prime\delta(z)$. This quantity, 
however, is vanishing as $A^\prime$ has a ${\rm sign}(z)$-like discontinuity
at $z=0$. We therefore obtain the following equation of motion for the
$(D-1)$-dimensional graviton components $H_{\mu\nu}$:
\begin{equation}\label{EOM1oGBXY}
 \Big(\Omega_{\mu\nu}+{\widehat M}_P^{3-D} 
 T_{\mu\nu}\Big) \delta(z)=0~.
\end{equation}
Note that this equation is purely $(D-1)$-dimensional. Thus, gravity is
{\em  completely} localized on the brane, that is at the $z=0$ hypersurface
$\Sigma$. In particular, the graviton field $H_{\mu\nu}$
is non-vanishing only on the brane, while it vanishes in the bulk:
\begin{equation}\label{Hbulk}
 H_{\mu\nu}(z\not=0)=0~.
\end{equation}
Note that (\ref{EOM1oGBXY}) does not by itself imply 
(\ref{Hbulk}). In particular, {\em a priori} $H_{\mu\nu}$ at $z\not=0$ can be
arbitrary. However, as we explained above, we have no propagating degrees of
freedom in the bulk, that is, the graviton propagator in the bulk vanishes, 
while it is non-vanishing only on the brane. This implies that perturbations
due to matter localized on the brane should {\em not} propagate into the bulk
but only on the brane, hence (\ref{Hbulk}).

{}On the brane (\ref{EOM1oGBXY}) can be solved in a standard way. From 
(\ref{EOM1oGBXY}) it is clear that ${\widehat M}_P$ is the 
$(D-1)$-dimensional Planck scale for $(D-1)$-dimensional 
gravity localized on the brane. Actually, 
${\widehat M}_P$ is identified with the $(D-1)$-dimensional Planck scale
for the positive $\Delta$ solution. As to the negative $\Delta$ solution,  
we have ``antigravity'' localized on the brane, and the corresponding
theory is non-unitary due to negative norm states propagating on the brane.

{}Note that above our analysis was confined to the linearized theory.
The above conclusions, however, are valid in the full non-linear theory.
Indeed, we have no propagating degrees of freedom in the bulk, while on the
brane we have only the zero mode for the $(D-1)$-dimensional graviton 
components $H_{\mu\nu}$. This then implies that in the solitonic brane
world background (the gravitational part of) the
brane world-volume theory is described by the
$(D-1)$-dimensional Einstein-Hilbert action:
\begin{equation}\label{actionWV}
 S_{\rm {\small brane}}={\widehat M}_P^{D-2}
 \int d^{D-1} x \sqrt{-{\widehat G}} 
 {\widehat R}~,
\end{equation}
where ${\widehat G}_{\mu\nu}$ is the $(D-1)$-dimensional 
metric on the brane; all the
hatted quantities are $(D-1)$-dimensional, and are constructed from 
${\widehat G}_{\mu\nu}$. Note that there is no 
$(D-1)$-dimensional Gauss-Bonnet term in this action, which can be seen by
examining the full non-linear 
equations of motion following from (\ref{actionGB}).

\section{Comments}

{}We would like to end our discussion with the following remarks. As in the 
case of a solitonic brane world solution of \cite{alberto} with a space-like
extra dimension, in the above solution with a time-like extra dimension there
exist consistent curved deformations (that is, solutions with non-vanishing 
brane cosmological constant on the brane). However, if we embed our solution in
the (minimal)
supergravity framework (such an embedding exists in complete parallel 
with the solution discussed in \cite{alberto}), then the corresponding 
solitonic brane world is a BPS solution which preserves 1/2 of the original
supersymmetries. The $(D-1)$-dimensional cosmological constant on such a 
BPS brane is then necessarily vanishing.

{}Let us point out that, just as is the case for the solitonic brane world 
solution of \cite{alberto}, the solution discussed in this paper does not 
suffer from the difficulties such as delocalization of gravity 
\cite{COSM,zuraRS,olindo} or
inconsistency of the coupling between brane matter and bulk gravity 
\cite{zuraRS,Lang}, which are generically expected to occur at the quantum 
level in warped backgrounds such as \cite{RS}\footnote{Here we should
point out that, as in the model of \cite{alberto}, we can construct
the above $\delta$-function-like solitonic solution as a limit of a 
thick solitonic domain wall. As was discussed in detail in \cite{alberto},
in the corresponding solution only the graviton zero mode is a propagating
solution.}. 

{}In the above solitonic solution perturbatively we have no propagating 
degrees of freedom in the bulk. At first this might appear to imply that 
the extra dimension is immaterial. Note, however, that, as was pointed out
in \cite{alberto}, non-perturbative corrections in the bulk can have
non-trivial implications. In particular, semi-classically causality can be 
broken via creation of ``baby branes'' (which, nonetheless, need not
violate unitarity even if we have a time-like extra dimension). An interesting
phenomenological implication of this would be violation of global quantum
numbers such as the baryon and lepton numbers along the lines of \cite{baby}.

{}Finally, in the above setup
the classical choice of parameters given by (\ref{fine}) is {\em necessary}
to ensure unitarity as the extra dimension is time-like,
so this choice is {\em not} a fine-tuning, but rather is required by
unitarity\footnote{Here we
would like to point out that, as was discussed in \cite{Zanelli}, 
the corresponding choice of parameters in the case of a space-like
extra dimension is {\em not} a fine-tuning either.}. 
On the other hand, (\ref{fine}) is stable against
loop corrections both in the setup of this paper as well as that of 
\cite{alberto}.  

\acknowledgments

{}This work was supported in part by the National Science Foundation.
Z.K. would also like to thank Albert and Ribena Yu for financial support.

\end{document}